\documentclass[10pt,conference]{IEEEtran}
\IEEEoverridecommandlockouts
\usepackage{cite}
\usepackage{amsmath,amssymb,amsfonts}
\usepackage{algorithm}
\usepackage{algpseudocode}
\usepackage{graphicx}
\usepackage{textcomp}
\usepackage{xcolor}
\usepackage{booktabs}
\usepackage{multirow}
\usepackage{subcaption}
\def\BibTeX{{\rm B\kern-.05em{\sc i\kern-.025em b}\kern-.08em
    T\kern-.1667em\lower.7ex\hbox{E}\kern-.125emX}}

\begin{document}

\title{Detecting Unauthorized Drones with Cell-Free Integrated Sensing and Communication\\

\author{\IEEEauthorblockN{Xinyue Li\IEEEauthorrefmark{1}, Zinat Behdad\IEEEauthorrefmark{1}, Ozan Alp Topal\IEEEauthorrefmark{1}, \"Ozlem Tu\u{g}fe Demir\IEEEauthorrefmark{2}, and Cicek Cavdar\IEEEauthorrefmark{1}} 
\IEEEauthorblockA{\IEEEauthorrefmark{1}Department of Computer Science, KTH Royal Institute of Technology, Stockholm, Sweden \\ {Email: \{xinyue2, zinatb, oatopal, cavdar\}@kth.se}}
\IEEEauthorblockA{\IEEEauthorrefmark{2}Department of Electrical-Electronics Engineering, TOBB ETU, Ankara, Türkiye  (ozlemtugfedemir@etu.edu.tr)}
	
}

%{\footnotesize \textsuperscript{*}Note: Sub-titles are not captured %in Xplore and
%should not be used}
\thanks{This work was supported by Swedish Innovation Agency Funded (VINNOVA) through the SweWIN center (2023-00572).
}
%\vspace{-4mm}
}
\maketitle
%\vspace{-4mm}

\begin{abstract}
Integrated sensing and communication (ISAC) boosts network efficiency by using existing resources for diverse sensing applications. In this work, we propose a cell-free massive MIMO (multiple-input multiple-output)-ISAC framework to detect unauthorized drones while simultaneously ensuring communication requirements. We develop a detector to identify passive aerial targets by analyzing signals from distributed access points (APs). In addition to the precision of the sensing, timeliness of the sensing information is also crucial due to the risk of drones leaving the area before the sensing procedure is finished. We introduce the age of sensing (AoS) and sensing coverage as our sensing performance metrics and propose a joint sensing blocklength and power optimization algorithm to minimize AoS and maximize sensing coverage while meeting communication requirements. Moreover, we propose an adaptive weight selection algorithm based on concave-convex procedure to balance the inherent trade-off between AoS and sensing coverage. Our numerical results show that increasing the communication requirements would significantly reduce both the sensing coverage and the timeliness of the sensing. Furthermore, the proposed adaptive weight selection algorithm can provide high sensing coverage and reduce the AoS by $45\%$ compared to the fixed weights, demonstrating efficient utilization of both power and sensing blocklength. 
%This study examines an integrated sensing and communication (ISAC) framework designed for detecting unauthorized unmanned aerial vehicles (drones) over a wide area. The ISAC transmit access points (APs) work collectively to serve user equipments (UEs) while potentially directing a beam towards the passive aerial target. A detector, employing a maximum a posteriori ratio test, is devised to pinpoint the target by analyzing signals received from distributed APs. We mainly focus on enhancing the sensing speed of the system by optimizing sensing block length and cooperation cluster size.
\end{abstract}
%%%%%%%%%%%%%%%%%%%%%%%%%%%%%%%%%%%%%%%%%%%%%%%%%%%%%%%%%%%
%\vspace{-2mm}
\section{Introduction}\label{section1}
Integrated sensing and communication (ISAC) has emerged as a promising paradigm in 6G and future wireless networks, enabling simultaneous communication and sensing within the same infrastructure. By leveraging ISAC, wireless systems can efficiently detect and track terrestrial and aerial targets while maintaining seamless communication services. Cell-free massive multiple-input multiple-output (MIMO) systems, characterized by distributed antennas, achieve higher spectrum efficiency than traditional small-cell setups supporting both terrestrial and aerial user equipments (UEs)\cite{cell-free-book, zheng2021uav}. The distributed nature of cell-free systems offers unique advantages for ISAC implementation by enhancing coverage, spatial diversity, and improved detection accuracy, making them well-suited for detecting unauthorized drones.
 
In the context of cell-free massive MIMO-ISAC systems, previous studies have primarily focused on terrestrial targets. The work in \cite{behdad2022power} introduces a maximum a posteriori ratio test detector and a power allocation strategy to enhance sensing performance. %Meanwhile, \cite{demirhan2023cell} focuses on boosting the sensing signal-to-noise ratio (SNR) using conjugate and null-space beamforming to optimize sensing without disrupting communication.
However, aerial targets such as drones have unique characteristics due to their altitude and distance from APs. The authors in \cite{Shivani2024uavdetection} propose a sensing-centric (SC) power allocation strategy that maximizes the detection SNR to enhance detection performance for an aerial target in orthogonal time frequency space
(OTFS)-aided cell-free massive MIMO ISAC systems.  

For target detection, two key metrics are detection and false alarm probabilities, reflecting the precision of sensing information. With aerial targets, the system must ensure high detection probability across a wide area. Thus, we define sensing coverage as the percentage of the area where detection probability exceeds a threshold. Alongside precision and coverage, the timeliness of sensing is crucial for detecting drones before they leave the area. The age of sensing (AoS) quantifies the freshness of sensing data, representing the time between the current moment and when the data was generated, as introduced in \cite{zheng2024average}. AoS depends on the time required to update sensing information.
In \cite{Jing2023Peak}, two schemes are proposed to minimize the peak age of information (PAoI) in the internet of vehicles (IoV) environments: the sensing-computing-computation (SCC) scheme, where vehicles handle all tasks, and the partial-offloading SCC (PO-SCC) scheme, where vehicles offload computational tasks to roadside units. These algorithms optimize beamforming, resource allocation, and offloading to reduce PAoI, but do not incorporate ISAC or cell-free massive MIMO systems.

%There is an inherent trade-off between sensing coverage and the AoS: reducing the total observation time decreases AoS but can lower detection probability.
In this paper, we define two key tasks for ISAC networks: the communication task, ensuring that users meet a minimum signal-to-interference-plus-noise ratio (SINR), and the sensing task, aimed at detecting unauthorized drones. Our main goal is to minimize AoS while maximizing sensing coverage, all while meeting communication requirements.
To minimize AoS, we reduce the observation period at each sensing location, which is influenced by the sensing blocklength. Since power is shared between sensing and communication, we propose a joint optimization algorithm that adjusts sensing blocklength and power to maximize sensing coverage and minimize AoS while ensuring communication performance. 
However, some locations may require longer observation times for high sensing coverage. The main contribution of this paper is to balance this trade-off by introducing a novel weight selection algorithm that dynamically adjusts the trade-off between sensing coverage and AoS. This algorithm efficiently allocates power and blocklength to maintain high detection probability while minimizing sensing time, ultimately improving network performance.
%\vspace{-2mm}
\enlargethispage{-0.1in}
\section{System Model}
\label{section3}
\begin{figure}[t]
\vspace{4mm}
\centerline{\includegraphics[trim={1mm 0mm 0mm 0mm},clip,width=0.6\linewidth]{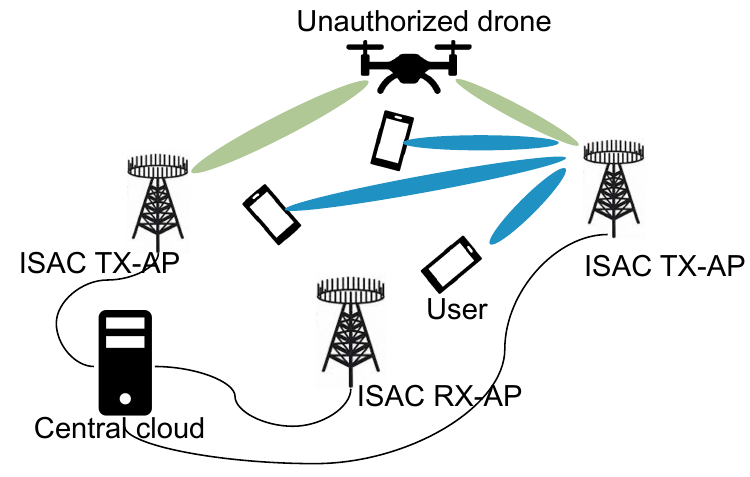}}
%\vspace{-3mm}
\caption{Illustration of the ISAC system setup.}
\label{fig1}
\vspace{-4mm}
\end{figure}

We consider an ISAC system in a cell-free massive MIMO setup on top of the centralized radio access network (C-RAN) architecture, as illustrated in Fig.~\ref{fig1}, featuring downlink communication and multi-static sensing, where sensing transmitters and receivers are not co-located. The system comprises $L$ terrestrial ISAC transmit APs and $R$ terrestrial sensing receive APs, all interconnected through fronthaul links to a central cloud and fully synchronized \cite{cell-free-book}. We define $\mathcal{L}=\{1, \cdots, L\}$ and $\mathcal{R}=\{1, \cdots, R\}$ as the set of transmit and receive APs, respectively. Each AP is equipped with $M$ antennas arranged in a horizontal uniform linear array (ULA).

All ISAC transmit APs jointly serve $K$ terrestrial UEs by transmitting centralized precoded signals. Simultaneously, the APs contribute to the detection of aerial targets, such as drones, by transmitting an additional sequence of $\tau_s$ sensing symbols towards candidate sensing locations. The sensing signal utilizes the same time-frequency resources as the communication signal, with $\tau_s$ denoting the sensing blocklength.
A beam-searching method is employed, whereby the transmit APs cooperatively direct a sensing beam towards a predetermined location during a single observation period. In parallel, the receive APs actively detect the presence of the target before shifting their focus to the subsequent location. All signal processing tasks are centralized in the cloud, with a line-of-sight (LOS) connection assumed between each AP and the target.

\subsection{Signal Model and Downlink Communication} 
We define the communication/sensing channel from all the $LM$ transmit antennas to UE $k$ and the target in the network as  $\textbf{h}_{k}^*=\begin{bmatrix}
\textbf{h}_{k,1}^H& \ldots&
\textbf{h}_{k,L}^H
\end{bmatrix}^T\in \mathbb{C}^{LM}$ and $\textbf{h}_{0}^*= \begin{bmatrix}
\textbf{a}^T(\varphi_{1},\vartheta_{1})&  \ldots&
\textbf{a}^T(\varphi_{L},\vartheta_{L})
\end{bmatrix}^T\in \mathbb{C}^{LM}$, respectively. 
Here, $\textbf{h}_{k,l}^*\in \mathbb{C}^M$ denote the channel from transmit AP $l$ to UE $k$ and $\textbf{a}(\varphi_{l},\vartheta_{l})\in \mathbb{C}^M$ is the antenna array response vector corresponding to transmit AP $l$ with $\varphi_{l}$ and $\vartheta_{l}$ as the azimuth and elevation angles from transmit AP $l$ to the target location, respectively.
Assuming the antennas at the APs are half-wavelength-spaced,  
$\textbf{a}(\varphi_l,\vartheta_l) =\begin{bmatrix}
          1& e^{j\pi \sin(\varphi_l)\cos(\vartheta_l)}& \ldots& e^{j(M-1)\pi\sin(\varphi_l)\cos(\vartheta_l)}
        \end{bmatrix} ^T$ \cite{bjornson2017massive}. 
\enlargethispage{-0.05in}
We define $s_k[m]$ as the zero-mean downlink communication symbol for UE $k$ at time instance $m$ with unit power, i.e., $\mathbb{E}\{|s_k[m]|^2\}=1$. The sensing symbol, $s_0[m]$, is fixed at 1. The transmitted signal $\textbf{x}_l[m] \in \mathbb{C}^M$ from transmit AP $l$ at time instance $m$ can be written as
\begin{equation}\label{x_k}%\vspace{-1mm}
    \textbf{x}_l[m]= \sum_{k=0}^{K} \sqrt{\rho_{k}}\textbf{w}_{k,l} s_{k}[m], 
    \quad l\in \mathcal{L}
   % \vspace{-1mm}
\end{equation} 
\enlargethispage{-0.1in}
where we use a common power control coefficient for each UE, $\rho_k\geq 0$, and for the target, $\rho_0\geq0$ and $\textbf{w}_{k,l}\in \mathbb{C}^{M}$ and $\textbf{w}_{0,l}\in \mathbb{C}^M$ are the transmit precoding vectors for transmit AP $l$ corresponding to UE $k$ and the sensing signal, respectively. The precoding vectors for each UE and the target are jointly selected based on the perfect channel state information (CSI) from all the $LM$ distributed transmit antennas. The average transmit power for transmit AP $l$ is computed as
%\vspace{-1.5mm}
\begin{equation} \label{eq:Pk}
    P_l = \mathbb{E}\{\Vert\textbf{x}_l[m]\Vert^2\} = \sum_{k=0}^{K}\rho_k\Vert \textbf{w}_{k,l} \Vert^2, \quad l\in \mathcal{L}.
    %\vspace{-1.5mm}
\end{equation}

The precoding vectors for each transmit AP are extracted from the concatenated centralized precoding vectors given as $
        \textbf{w}_k = \begin{bmatrix}
\textbf{w}_{k,1}^T & \textbf{w}_{k,2}^T & \hdots & \textbf{w}_{k,L}^T
\end{bmatrix}^T\in \mathbb{C}^{LM}$,
for $k=0,1,\ldots,K$. Using the common coefficients and centralized precoding approach ensure interference control based on the overall channel from the $LM$ antennas.

%\subsection{Transmit Precoding Vectors}
%\vspace{-0.5mm}
%Precoding vectors for UEs and the target are jointly determined using channel state information (CSI) from all $LM$ distributed transmit antennas.
The unit-norm regularized zero-forcing (RZF) precoding vector is  constructed for UE $k$ as $\textbf{w}_{k}=\frac{\bar{\textbf{w}}_{k}}{\left \Vert \bar{\textbf{w}}_{k}\right \Vert}$ \cite{cell-free-book}, where
%\vspace{-3mm}
\begin{equation}
     \bar{\textbf{w}}_{k}
     \!=\!\left(\sum\limits_{j=1}^{K}\textbf{h}_j\textbf{h}_j^H+\lambda\textbf{I}_{LM}\right)^{\!\!-1}\!\!\textbf{h}_k, \ k=1,\ldots,K,
     \vspace{-2mm}
\end{equation}
and $\lambda$ is the regularization parameter. %Note that since the communication symbols also contribute to sensing by the reflected paths towards the target, it is not considered as interference for the sensing target. Hence, we aim at nulling the interference only for the UEs. To null the destructive interference from the sensing signal to the UEs, the sensing precoding vector $\textbf{w}_{0}$ can be selected as the ZF precoder, i.e., by projecting $\textbf{h}_0$ onto the nullspace of the subspace spanned by the UE channel vectors as in \cite{buzzi2019using}.%, i.e., 
For sensing signals, we use unit-norm maximum ratio transmission (MRT) precoding approach, as $\textbf{w}_0 = \frac{\textbf{h}_0}{\Vert \textbf{h}_0\Vert}$.
%\subsection{Downlink Communication}
The received signal at UE $k$ is 
\vspace{-1mm}
\begin{align}    \label{y_i}
     z_k[m] %=&\sum_{l=1}^{L} \textbf{h}^{H}_{k,l}\textbf{x}_{l}[m]+ n_k[m]\nonumber \\
     =&\underbrace{\sqrt{\rho_k}\textbf{h}_{k}^{H}\textbf{w}_{k} s_{k}[m]}_{\textrm{Desired signal}}+ \underbrace{\sum_{j=1,j\neq k}^{K}\sqrt{\rho_j}\textbf{h}_{k}^{H}\textbf{w}_{j} s_{j}[m]}_{\textrm{Interference signal due to the other UEs}}\nonumber\\
    &+ \underbrace{\sqrt{\rho_0}\textbf{h}_{k}^{H}\textbf{w}_{0} s_{0}[m]}_{\textrm{Interference signal due to the sensing}}+ \underbrace{n_k[m]}_{\textrm{Noise}} ,
\end{align}
where the thermal noise at the receiver of UE $k$  is represented by $n_k[m] \sim \mathcal{CN}(0,\sigma_n^2)$. The SINR at UE $k$ is given by

\begin{align}\label{sinr_i}
     \gamma_k 
     &= \frac{\rho_k\left\vert \textbf{h}_k^H \textbf{w}_k\right\vert^2}{\sum_{j=1,j\neq k}^{K}\rho_j\left\vert \textbf{h}_k^H \textbf{w}_j\right\vert^2+\rho_0\left\vert \textbf{h}_k^H \textbf{w}_0\right\vert^2+\sigma_n^2}.
\end{align}

\subsection{Target Detection}
We consider multi-static sensing, meaning that the sensing transmitters and the receivers are not co-located. We assume that the target-free channel between transmit AP $l$ and receive AP $r$ is acquired prior to sensing in the absence of the target. The transmit signal $\textbf{x}_l[m]$ is also known at the central cloud. Hence, except the noise, the undesired part of the received signal at each receive AP can be cancelled.
According to $s_0[m]=1$, the received signal at AP $r\in\mathcal{R}$ in the presence of the target can be expressed as\footnote{In practice, one should also take the cancellation error into account, which is left as future work.}
\enlargethispage{-0.1in}
\vspace{-2mm}
\begin{equation}\label{y_r}
          \textbf{y}_r[m]  = \sum_{l=1}^{L} \alpha_{r,l} 
          \sqrt{\beta_{r,l}}\textbf{a}(\phi_{r},\theta_{r})\textbf{a}^{T}(\varphi_{l},\vartheta_l) \sqrt{\rho_{0}}\textbf{w}_{0,l}+\textbf{n}_r[m],\nonumber
          \vspace{-1mm}
\end{equation}
where $\textbf{n}_r[m]\sim \mathcal{CN}(\textbf{0},\sigma_n^2\textbf{I}_M)$ is the receiver noise at the $M$ antennas of receive AP $r$. The matrix $\alpha_{r,l}\, \sqrt{\beta_{r,l}}\,\textbf{a}(\phi_{r},\theta_{r})\,\textbf{a}^{T}(\varphi_{l},\vartheta_l)$ represents the reflected path through the target where $\phi_r$ and $\theta_r$ are the azimuth and elevation angles from the target location to receive AP $r$. Here,  $\beta_{r,l}$ is the combined sensing channel gain that includes the combined effect of the  path-loss of the path through the target and the radar cross section (RCS) variance of the target and $\alpha_{r,l}\sim \mathcal{CN}(0,\,1)$ represents the variation of RCS values, following the Swerling-I model, in which the velocity of the target is low compared to the total sensing duration. Hence, $\alpha_{r,l}$ is constant throughout the consecutive symbols collected for sensing. Moreover, we assume that they are independent for different APs. In accordance with the previous literature \cite{zhao2022joint}, we neglect the paths resulting from multi-reflections from the other objects due to the presence of the target. Moreover, reflections due to communication signals are disregarded, as they are primarily directed toward terrestrial UEs, making their impact on aerial target negligible.

For notational simplicity, we define
$\boldsymbol{\beta}_{r} \triangleq \begin{bmatrix}\sqrt{\beta_{r,1}} \textbf{a}^{T}(\varphi_{1},\vartheta_1)\textbf{w}_{0,1}&\dots& \sqrt{\beta_{r,L}}  \textbf{a}^{T}(\varphi_{L},\vartheta_L)\textbf{w}_{0,L}\end{bmatrix} ^T\in \mathbb{C}^{L}$
and $\boldsymbol{\alpha}_{r} \triangleq  \begin{bmatrix}
\alpha_{r,1}&\alpha_{r,2}&\ldots&\alpha_{r,L}\end{bmatrix}^T\in \mathbb{C}^{L}$. 
We use distributed maximum ratio combining (MRC) approach at each receive AP $r\in\mathcal{R}$ by multiplying the combining vector $\boldsymbol{v}_{r}^H=\frac{\textbf{a}^H(\phi_{r},\theta_{r})}{\|\textbf{a}(\phi_{r},\theta_{r})\|}$ with the received signal, resulting in
\vspace{-2mm}
\begin{align}\label{y_rPrim}
          y_r[m]  %&= \sqrt{M}\sqrt{\rho_{0}}\sum_{l=1}^{L} \alpha_{r,l}  \sqrt{\beta_{r,l}} \textbf{a}^{T}(\varphi_{l},\vartheta_l) \textbf{w}_{0,l}+n'_r[m]\nonumber\\
          &=\sqrt{M\rho_{0}}\,  \boldsymbol{\beta}_{r}^T\,\boldsymbol{\alpha}_{r} +n_r'[m],
          %\vspace{-1mm}
\end{align}
where $n_r'[m]=\frac{\textbf{a}^H(\phi_{r},\theta_{r})}{\|\textbf{a}(\phi_{r},\theta_{r})\|}\textbf{n}_r[m]\sim \mathcal{CN}(0,\sigma_n^2) $. Finally, we form the concatenated received signal $\textbf{y}[m]\in \mathbb{C}^{R}$ by all $R$ receive APs involved in the sensing, i.e., $r\in\mathcal{R}$, as follows
%\vspace{-1mm}
\begin{align} 
\label{y_rPrim}
          \textbf{y}[m] %&\triangleq \begin{bmatrix}y_1[m]\\\vdots\\y_{R}[m]\end{bmatrix} = \sqrt{M\rho_{0}}  \begin{bmatrix}\boldsymbol{\beta}_{1}^{T} \boldsymbol{\alpha}_{1}\\\vdots\\\boldsymbol{\beta}_{R}^{T} \boldsymbol{\alpha}_{R}\end{bmatrix}  + \textbf{n}[m]\nonumber\\
          &=  \sqrt{M\rho_{0}} \boldsymbol{\beta}^T \boldsymbol{\alpha} + \textbf{n}[m],
          %\vspace{-1mm}
\end{align}
where $\boldsymbol{\beta}^T= \operatorname{bdiag}\left(\boldsymbol{\beta}_1^T,\ldots, \boldsymbol{\beta}_{R}^T\right)\in \mathbb{C}^{R\times R\,L}$, $\boldsymbol{\alpha}= \begin{bmatrix}
    \boldsymbol{\alpha}_1^T & \ldots & \boldsymbol{\alpha}_{R}^T
\end{bmatrix}^T\in\mathbb{C}^{R\,L}$, and $\textbf{n}[m] = \begin{bmatrix}
    n_1'[m]& \ldots& n_{R}'[m]
\end{bmatrix}^T \in \mathbb{C}^{R}$.
%Finally, we define $\textbf{y}_{\tau}$ as the concatenated vector containg all the received signals $\textbf{y}[m]$ over $m=1, \ldots,\tau_s$, as follows 
%\begin{equation}\label{y_rPrim}
%          \textbf {y}_{\tau} \triangleq \underbrace{\textbf{1}_{\tau\times 1}\otimes (\sqrt{M\rho_{0}} \boldsymbol{\beta}^{T} \boldsymbol{\alpha})}_{ \textrm{Desired signal}}+ 
%    \underbrace{\begin{bmatrix}\textbf{n}_[1]\\\vdots\\\textbf{n}_[\tau]\\\end{bmatrix} }_{ \textrm{Noise}}       
%\end{equation}

To detect the target we have two hypotheses as follows:
\begin{align}
    &\mathcal{H}_0 : \textbf{y}[m]= \textbf{n}[m],& \quad m=1,\ldots,\tau_s,
    \\
    &\mathcal{H}_1 : \textbf{y}[m]= \sqrt{M\rho_{0}} \boldsymbol{\beta}^T \boldsymbol{\alpha}+\textbf{n}[m],&\quad m=1,\ldots,\tau_s,
\end{align}
where  $\mathcal{H}_0 $ represents the hypothesis that there is no target and   $\mathcal{H}_1 $ represents the hypothesis that the target exists and the reflected signals from the target are received by the receive APs.
We employ a maximum a posteriori ratio test (MAPRT) detector with the derived test statistic 
\begin{align}
    T = \left(\sum_{m=1}^{\tau_s}\textbf{y}^H[m]\right) \mathbf{B} \left(\sum_{m=1}^{\tau_s}\textbf{y}[m]\right)\label{eq:MAPRT}
\end{align}
where 
\begin{align}\label{eq:B}
    \mathbf{B} = M \rho_0 \boldsymbol{\beta}^T \left(M\rho_0\tau_s \boldsymbol{\beta}^* \boldsymbol{\beta}^T + \sigma_n^2 \textbf{I}_{RL}\right)^{-1} \boldsymbol{\beta}^*.   \end{align}
%The proof is presented in Appendix A.

We define the detector threshold as $\lambda$, determined for a given false alarm probability. Finally, the true hypothesis $\hat{\mathcal{H}}$ is estimated as 
\begin{align}
    \hat{\mathcal{H}} = \left\{\begin{matrix}
\mathcal{H}_0, & \textrm{if}\quad T<\lambda, \\
\mathcal{H}_1, & \textrm{if} \quad T\geq\lambda. \\
\end{matrix}\right. 
\end{align}
The expectation of test statistics under hypothesis $\mathcal{H}_0$ and $\mathcal{H}_1$ are given as
\begin{align}
    &\mathbb{E}\{T|\mathcal{H}_0\}= \tau_s\,\mathbb{E}\left\{\textbf{n}^H[m]\, \mathbf{B} \,\textbf{n}[m])\right\}=\tau_s\, \sigma_n^2 \operatorname{tr}(\mathbf{B})\\
    %&=\tau_s\, \sigma_n^2 \operatorname{tr}(\textbf{B})\nonumber\\
    &\hspace{7mm} = \tau_s\,\sigma_n^2 M\,\rho_0 \operatorname{tr}\left(\boldsymbol{\beta}^*\boldsymbol{\beta}^T \left(M\rho_0\tau_s \boldsymbol{\beta}^* \boldsymbol{\beta}^T + \sigma_n^2 \textbf{I}_{RL}\right)^{-1} \right),\nonumber\\
    &\mathbb{E}\{T|\mathcal{H}_1\}= \tau_s^2 M \rho_0 \,\mathbb{E}\left\{\boldsymbol{\alpha}^H \boldsymbol{\beta}^* \,\mathbf{B}\, \boldsymbol{\beta}^T \,\boldsymbol{\alpha}\right\} \label{eq:T_H1}\\
    &\hspace{5mm}+\tau_s\,\mathbb{E}\left\{\textbf{n}^H[m]\, \mathbf{B} \,\textbf{n}[m])\right\},\nonumber\\
    &= (\tau_s M \rho_0)^2 \operatorname{tr} \left(\boldsymbol{\beta}^*\boldsymbol{\beta}^T \left(M\rho_0\tau_s \boldsymbol{\beta}^* \boldsymbol{\beta}^T + \sigma_n^2 \textbf{I}_{RL}\right)^{-1}\boldsymbol{\beta}^*\boldsymbol{\beta}^T\right)\nonumber\\
    &\hspace{5mm} + \tau_s\, \sigma_n^2 M\,\rho_0 \operatorname{tr}\left(\boldsymbol{\beta}^*\boldsymbol{\beta}^T \left(M\rho_0\tau_s \boldsymbol{\beta}^* \boldsymbol{\beta}^T + \sigma_n^2 \textbf{I}_{RL}\right)^{-1} \right)\nonumber,
\end{align}
respectively, where we have used the cyclic shift property of the trace operation and inserted $\mathbf{B}$ in obtaining \eqref{eq:T_H1} and used $\mathbb{E}\{\boldsymbol{\alpha} \,\boldsymbol{\alpha}^H\}=\textbf{I}_{RL}$.
%\begin{align}
  %  \mathbb{E}\{T|\mathcal{H}_1\}%&= (\tau_s)^2 M \rho_0\,\mathbb{E} \{ \operatorname{tr}(\boldsymbol{\alpha}^H\, \boldsymbol{\beta}^* \,\mathbf{B} \,\boldsymbol{\beta}^T \,\boldsymbol{\alpha})\} +\tau_s\, \sigma_n^2 \operatorname{tr}(\textbf{B})\nonumber\\
    %&=\tau_s^2 M \rho_0 \mathbb{E} \{ \operatorname{tr} ( \boldsymbol{\beta}^* \,\mathbf{B}\, \boldsymbol{\beta}^T\, \boldsymbol{\alpha} \,\boldsymbol{\alpha}^H ) \} +\tau_s\, \sigma_n^2 \operatorname{tr}(\textbf{B})\nonumber\\
    %&=\tau_s^2 M \rho_0 \, \operatorname{tr} \left( \boldsymbol{\beta}^* \mathbf{B} \boldsymbol{\beta}^T \underbrace{\mathbb{E} \{ \boldsymbol{\alpha} \boldsymbol{\alpha}^H \}}_{\textbf{I}} \right)+\tau_s\, \sigma_n^2 \operatorname{tr}(\textbf{B})\nonumber\\
  %  &=\tau_s^2 M \rho_0 \operatorname{tr} \left( \boldsymbol{\beta}^* \mathbf{B} \boldsymbol{\beta}^T\right)+\tau_s\, \sigma_n^2 \operatorname{tr}(\textbf{B})
%\end{align}
%\vspace{2mm}
\enlargethispage{-0.1in}
\section{Drone Detection with ISAC}
As discussed in the previous section, we consider two main goals for designing an ISAC system for unauthorized drone detection: (i) high precision, and (ii) freshness of the decision. Aiming the wide area drone detection, we consider sensing coverage as the main metric modeling the precision, while the age of sensing models the freshness of the decision. Below, we detail our sensing performance metrics.
\subsection{Age of Sensing}
The freshness of the decision is important especially for drones. Due to their high speed, a taken decision should be refreshed with a high speed. For example, if the drone moves faster than the sensing duration of the whole map, it can be completely undetected by the ISAC system. This brings the necessity to model the freshness of the sensing decision. 
Inspired by \cite{zheng2024average}, we define the AoS as the elapsed time since the latest target existence decision at a specific location. We express AoS at time $t$ as:
$\Delta(t) = t- U_u(t)$,
where $U_u(t)$ marks the time of the latest decision at that specific location. Here, the subscript $u$ represents the index of the latest update. We consider a wide-area sensing scenario in which the proposed system scans $S$ equidistant points across the area. Our system scans distributed points sequentially, making the maximum AoS the interval until all points are checked—equivalent to the total observation time. Since in the considered procedure, area and locations are predetermined, the total AoS of the system can be modeled deterministically as
\begin{align}
\vspace{-2mm}
\Delta_{\mathrm{total}} = \sum_{s=1}^{S} T_{s}(\tau_s)\approx  \sum_{s=1}^{S} \frac{\tau_s }{B},%\vspace{-4mm}
\end{align}
where $T_s$ is the time to observe location $s$, depending on the sensing blocklength $\tau_s$. Since the major influencing factor for the AoS is sensing blocklength $\tau_s$, we can approximate AoS as the right term above, where $B$ is the symbol rate. %Different sensing requirements may extend observation times at certain locations.
% so the worst-case observation time is considered when multiple locations are sensed simultaneously.
%\vspace{-1mm}
\subsection{Sensing Coverage}
We let $P_{\text{d}}(s)$ denote the detection probability in the sensing location $s$ and $P_{\text{th}}$ be the desired detection probability threshold. Given that the system senses $S$ sensing locations, we define the sensing coverage as the percentage that the system can achieve detection probability over the threshold, given by
\vspace{-2mm}
\begin{equation}
    A_c= \frac{\sum_{s=1}^{S} u(P_{\text{d}}(s) - P_{\rm th})}{S}\times 100, \vspace{-2mm}
\end{equation} 
where $u(x_s)$ is the unit step function, the value of which is equal to one if $x_s\geq 0$ and zero otherwise. 

\enlargethispage{-0.1in}
\subsection{Multi-Objective Optimization Problem}
An ideal ISAC system for wide area drone sensing aims maximizing the sensing coverage while minimizing the AoS subject to the communication constraints. These two goals conflict with each other since higher sensing duration increases the detection probability while also increasing the AoS. Therefore, we cast the ISAC wide area drone detection problem as a multi-objective problem as below:
\vspace{-2mm}
\begin{subequations}
  \begin{align}
     \underset{ \tau_{\text{s}}, \{\rho_k\}_{k=0}^K}{\text{maximize}} \quad&  \omega_0 \left( \frac{\sum_{s=1}^{S} u(P_{\text{d}}(s) - P_{\rm th})}{S} \right)- \omega_1 \sum_{s=1}^{S} \frac{\tau_s }{B}\label{eq:optimization0:objective}\\
     \text{subject to} \quad & \gamma_k \geq  \gamma_{\rm c},\quad k=1,\ldots,K \label{eq:optimization0:SINRonstraint} \\
    & P_{l} \leq \rho_{\text{max}}, \quad  l=1,\ldots, L \label{eq:optimization:powerconstraint}\\
    & \tau_{\text{s,min}} \leq \tau_{\text{s}} \leq \tau_{\text{s,max}}, \quad s=1,\ldots, S,   \label{eq:optimization:taus_constraint}
  \end{align}
\end{subequations}
where $\omega_0$ and $\omega_1$ are the weights for the sensing coverage and the AoS, respectively. As $w_0$ increases, the problem is more focused on maximizing the sensing coverage, while as $w_0$ gets close to 0, the problem will minimize the AoS. $\gamma_c$ is the SINR constraint that maintains the communication link, which means if the total downlink power is fixed, the higher $\gamma_c$ is, the lower the sensing power will be allocated. $\rho_{\text{max}}$ is the per-AP power budget.

 {\textit{\textbf{Remark 1:} Given that there is no resource sharing between sensing points, maximizing the sensing coverage is equivalent to maximizing the detection probability at each sensing point  $s$. Similarly, minimizing AoS corresponds to minimizing the sensing duration at each point. Therefore, the optimization problem can be addressed independently for each of the $S$ sensing points, where $\tau^*_s$ and $P^*_d(s)$ are the optimal sensing blocklength and detection probability for the sensing point $s$. In this case, the optimal AoS and sensing coverage can be respectively given as $\Delta^*_{\mathrm{total}} = \sum_{s=1}^{S} \frac{\tau^*_s }{B}$, and $A^*_c = \frac{\sum_{s=1}^{S} u(P^*_{\text{d}}(s) - P_{\rm th})}{S}$  }}.
\enlargethispage{-0.1in}

 Due to the analytical intractability of the detection probability and false alarm probability, we will utilize the test statistic, $\mathbb{E}\{T|\mathcal{H}_1\}-\mathbb{E}\{T|\mathcal{H}_0\}$, as a measure of precision. In this case, for an arbitrary sensing point $s$, the multi-objective optimization problem can be given as 
\begin{subequations}\label{eq:optimization-18}
  \begin{align}
     \underset{ \tau_{\text{s}}, \{\rho_k\}_{k=0}^K}{\text{maximize}}&\quad  \omega_0 \left(\mathbb{E}\{T|\mathcal{H}_1\}-\mathbb{E}\{T|\mathcal{H}_0\}\right)- \omega_1  \tau_{\text{s}} \label{eq:optimization0:objective}\\
     \text{subject to} &\quad \gamma_k \geq  \gamma_{\rm c},\quad\quad k=1,\ldots,K \label{eq:optimization0:SINRonstraint} \\
    & \quad P_{l} \leq \rho_{\text{max}}, \quad\quad l=1,\ldots, L \label{eq:optimization:powerconstraint}\\
    & \quad \tau_{\text{s,min}} \leq \tau_{\text{s}} \leq \tau_{\text{s,max}}. \label{eq:optimization:taus_constraint}
  \end{align}
\end{subequations}
The first objective is to maximize $\mathbb{E}\{T|\mathcal{H}_1\}-\mathbb{E}\{T|\mathcal{H}_0\}$ and the second objective is to minimize the sensing blocklength, $ \tau_{\text{s}}$. We first define a new optimization variable $\overline{\rho}_k \triangleq \rho_k \tau_s$. Expanding (\ref{eq:optimization0:SINRonstraint}), we get
\vspace{-2mm}
\begin{align}
    \sum_{j=1, j\neq k}^{K} \overline{\rho}_j \vert \textbf{h}_k^H \textbf{w}_j\vert ^2 + \overline{\rho}_0  \vert \textbf{h}_k^H \textbf{w}_0\vert ^2 + \sigma_n^2\tau_{\textrm{s}} \leq \overline{\rho}_k \frac{\vert \textbf{h}_k^H \textbf{w}_k\vert ^2}{\gamma_c} \label{eq:gamma_constraint}
\end{align}
and expanding (\ref{eq:optimization:powerconstraint}), we get
\vspace{-2mm}
\begin{align}
    \sum_{k=0}^{K}\overline{\rho}_k\Vert \textbf{w}_{k,l} \Vert^2\leq \rho_{\text{max}}\tau_{\textrm{s}}, \quad l=1,\ldots,L.\label{eq:poweRonstraint}
\end{align}
\enlargethispage{-0.1in}
We also denote the eigenvalue decomposition of $\boldsymbol{\beta}^*\boldsymbol{\beta}^T$ by $\boldsymbol{\beta}^*\boldsymbol{\beta}^T=\mathbf{U}\mathbf{D}\mathbf{U}^H$ where $\mathbf{D}$ is the diagonal matrix with real-valued non-negative eigenvalues in descending order along its diagonal. We denote the $i$th eigenvalue by $d_i$. Then, the first term in the objective function can be written as
\begin{align}
    &\mathbb{E}\{T|\mathcal{H}_1\}-\mathbb{E}\{T|\mathcal{H}_0\} = (M \tau_s\rho_0)^2 \nonumber\\
    &  \times  \operatorname{tr} \left(\mathbf{U}\mathbf{D}\mathbf{U}^H \left(M\rho_0\tau_s \mathbf{U}\mathbf{D}\mathbf{U}^H + \sigma_n^2 \mathbf{U}\mathbf{U}^H\right)^{-1}\mathbf{U}\mathbf{D}\mathbf{U}^H\right)\nonumber\\
    %&=\tau_s^2 (M \rho_0)^2 \operatorname{tr} \left(\mathbf{D}\mathbf{U}^H\mathbf{U}\mathbf{D}\mathbf{U}^H\mathbf{U}\left(M\rho_0\tau_s\mathbf{D}+\sigma_n^2\mathbf{I}\right)^{-1}\mathbf{U}^H\mathbf{U}\right) \nonumber\\
    %& + \sigma_n^2\tau_s\, M\,\rho_0 \operatorname{tr}\left(\mathbf{U}\mathbf{D}\mathbf{U}^H\mathbf{U} \left(M\rho_0\tau_s \mathbf{D} + \sigma_n^2 \mathbf{I}\right)^{-1}\mathbf{U}^H \right) \nonumber \\
    %&=\tau_s^2 (M \rho_0)^2 \operatorname{tr} \left(\mathbf{D}^2\left(M\rho_0\tau_s\mathbf{D}+\sigma_n^2\mathbf{I}\right)^{-1}\right) \nonumber\\
    %& +\sigma_n^2 \tau_s\, M\,\rho_0 \operatorname{tr}\left(\mathbf{D} \left(M\rho_0\tau_s \mathbf{D} + \sigma_n^2 \mathbf{I}\right)^{-1} \right) \nonumber\\
    & = (M \tau_s \rho_0)^2\sum_{i=1}^{R} \frac{d_i^2}{M\rho_0\tau_sd_i+\sigma_n^2}
\end{align}
Defining $x_i=M\rho_0\tau_sd_i$, the objective function becomes
\vspace{-2mm}
\begin{equation}
\sum_{i=1}^{R} \frac{x_i^2}{x_i+\sigma_n^2}.\vspace{-3mm}
\end{equation}
To solve the optimization problem, we define $\frac{x_i^2}{x_i+\sigma_n^2}\geq y_i$ and maximize $\sum_{i=1}^{R}y_i$, instead, with the constraints
\vspace{-2mm}
\begin{align}
    x_i^2\geq y_ix_i+y_i\sigma_n^2, \quad  \forall i 
\end{align}
%which is equivalent to 
%\begin{align}
%    2x_i^2+x_i^2+y_i^2 \geq 2y_ix_i+x_i^2+y_i^2+2y_i\sigma_n^2-2x_i
%\end{align}
which is equivalent to
\vspace{-2mm}
\begin{align}
    3x_i^2+y_i^2\geq (x_i+y_i)^2+2y_i\sigma_n^2 ,\quad  \forall i \label{eq:constraint_yi}.
\end{align}
This constraint is non-convex. % because of the second order combinations of $x_i$ and $y_i$ is greater than or equal to some affine combination of $x_i$ and $y_i$.  % As a result, we use CCP them we have the lower bounded as 
%\begin{align}
 %  f\left(x_i,y_i\right)&\geq f\left(x_i^{(c-1)}, y_i^{(c-1)}\right)+ \left(\begin{bmatrix}
 %x_i \\ y_i
 %\end{bmatrix}-\begin{bmatrix}
 %x_i^{(c-1)} \\ y_i^{(c-1)}
 %\end{bmatrix}\right)^T\nonumber\\
 %&\times \nabla f\left ( x_i^{(c-1)},y_i^{(c-1)} \right ) 
 %\end{align}
Applying convex-concave procedure (CCP) to the left-hand side of \eqref{eq:constraint_yi}, we can linearize the second-order terms as 
%\begin{align}
%    3x_i^2 + y_i^2 \geq 6\, x_i^{(c-1)} x_i + 2\, y_i^{(c-1)}y_i -3\,\left(x_i^{(c-1)}\right)^2 - \left(y_i^{(c-1)}\right)^2
%\end{align}
%Thus, the last constraint can be written as

\begin{align}
    6\, x_i^{(c-1)} &x_i + 2\, y_i^{(c-1)}y_i -3\,\left(x_i^{(c-1)}\right)^2 - \left(y_i^{(c-1)}\right)^2  \nonumber\\
    & \geq (x_i+y_i)^2+2y_i\sigma_n^2, \quad  \forall i, \label{eq:CCP-FPP-constraint}
\end{align}
where $c$ is the iteration index. CCP helps to transform the constraint into a quadratic convex constraint. The feasible point pursuit (FPP) approach can also be employed to guarantee a feasible start \cite{fpp}. The optimization problem can be given as \begin{subequations}\vspace{-2mm}
\begin{align}
  &  \underset{ \tau_{\text{s}}, \{y_i, x_i\}_{i=1}^R , \{\overline{\rho}_k\}_{k=0}^K}{\text{maximize}}\quad  \omega_0 \sum_{i=1}^{R} y_i- \omega_1  \tau_{\text{s}} \label{eq:optimization0:objective2}\\
  & \text{subject to} \quad \eqref{eq:gamma_constraint}, \eqref{eq:poweRonstraint}, \eqref{eq:optimization:taus_constraint} , \eqref{eq:CCP-FPP-constraint}\nonumber\\
     &\hspace{17mm} x_i = Md_i\overline{\rho}_0, \quad \forall i, \label{eq:optimization:equation x}
\end{align} \label{eq:optimization:final}
\end{subequations}
where it is in a convex form and can be solved iteratively by any convex program solver until the convergence is satisfied. Note that, the solution of \eqref{eq:optimization:final} is not the global optimal of the original problem due to the CCP relaxation, however, it is guaranteed to be a Karush-Kuhn-Tucker (KKT) point if the original problem is feasible \cite{fpp}. 
\vspace{-1mm}
\enlargethispage{-0.1in}
\subsection{Adaptive Weight Selection Algorithm}
In the optimization problem \eqref{eq:optimization-18}, the values of $\omega_0$ and $\omega_1$ dictate the trade-off between precision and speed in sensing. Based on the chosen sensing point, assigning 
the same weights can give different detection probabilities and sensing blocklengths. To ensure high coverage while minimizing the AoS, we propose an iterative algorithm that selects the optimal weights for each sensing location. This results in a minimized sensing blocklength while ensuring the detection probability meets a minimum threshold.
The steps of the algorithm are outlined in Algorithm~\ref{algo:adaptive_weight_selection}. The algorithm starts by assigning the full weight to the precision, trying to utilize all sensing blocklength to satisfy the aimed detection probability. If it is not satisfied, that point is assumed to be under outage. If the threshold is satisfied, the algorithm gradually improves the weight of the AoS minimization until the detection probability is not satisfied. 
%The process of the algorithm is: We first select an appropriate initial weight $w_{0init}$ which is small enough to ensure a minimal sensing blocklength. Then the whole sensing process is completed and the detection probability is calculated. Then we compare it with $P_{\text{th}}$. If the detection probability is bigger than threshold, we break the iteration and go on with the next location. If detection probability of current iteration is smaller than threshold, we gradually increase $w_0$ and repeat the previous process until detection probability meets the threshold or $w_0$ reaches 1.
It is worth noting that, after going through all locations in the map, there might be some points whose detection probability cannot reach the threshold even using maximum blocklength. %such as corner location. These points are defined not covered. Afterwards, we can calculate $\Delta_{total}$ and changing several input parameters and threshold and compare different scenarios.

\begin{algorithm}[h]
	\caption{Adaptive weight selection for one sensing location} \label{algo:adaptive_weight_selection}
	\begin{algorithmic}[1]
		\State {\bf Initialization:} Set initial weight $w_0=0$, $w_1=1$, step size $r$ and iteration count $\zeta=0\leq \zeta_{\rm max}$
        \State $\zeta=\zeta+1$, $w_0=1-r*(\zeta_{\rm max}-\zeta)$ and $w_1=1-w_0$
		 \State Using CCP to solve  \eqref{eq:optimization:final} and obtain optimized blocklength $\tau_{s}$ and power coefficients $\rho_k$, for $k=0,1, \cdots, K$.
		\State Calculate test statistic $T$ from \eqref{eq:MAPRT} and obtain  detection probability $P_{\text{d}}$ using Monte Carlo trials.
        \State If $P_{\text{d}}>P_{\text{th}}$ or $\zeta=\zeta_{\rm max}$ , proceed to Step 6. Otherwise, return to Step 2.
		\State {\bf Output:} Optimal weights $\omega_0, \omega_1$, blocklength $\tau_s$, power coefficients $\rho_k$ and detection probability $P_{\text{d}}$.
	    \end{algorithmic}
\end{algorithm}

% \begin{align}
%   &  \underset{ \tau_{\text{s}}, y_i, x_i , \overline{\rho}_k}{\text{maximize}}\quad  \omega_0 \sum_{i=1}^{R} y_i- \omega_1  \tau_{\text{s}}  + \lambda \sum_{i=1}^{R} \chi_i\label{eq:optimization0:objective}\\
%   & \text{subject to} \quad \eqref{eq:gamma_constraint}, \eqref{eq:poweRonstraint},  \eqref{eq:CCP-FPP-constraint} \nonumber \\
%      & \hspace{17mm}\tau_{\text{s,min}} \leq \tau_{\text{s}} \leq \tau_{\text{s,max}}\\ 
%      &\hspace{17mm} x_i = Md_i\overline{\rho}_0, \quad \forall i  
% \end{align}

\enlargethispage{-0.1in}
\section{Numerical Analysis}
\begin{figure}[t]
\vspace{0mm}
    \centering
    \includegraphics[trim={0mm 0mm 0mm 8mm},clip,width=0.8\columnwidth]{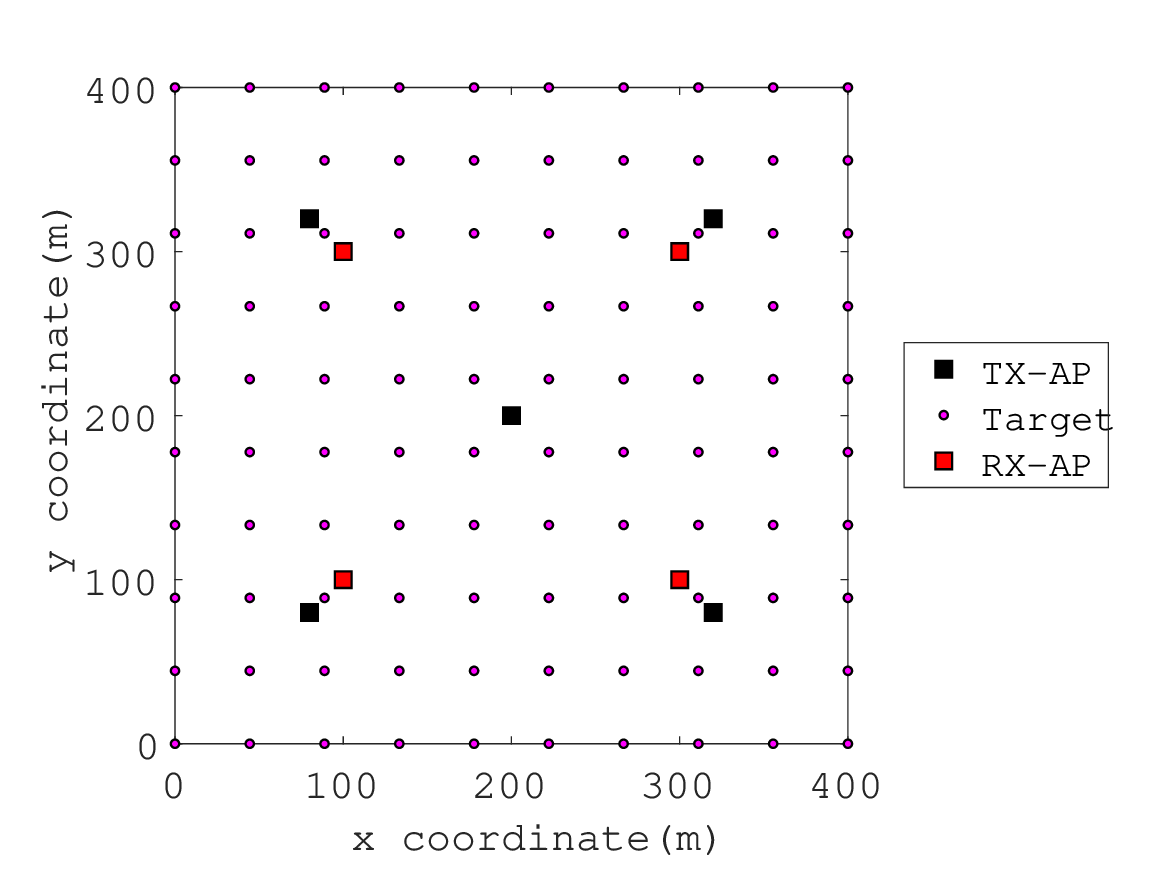}
    \vspace{-2mm}
    \caption{Locations of sensing points, TX APs and RX APs for the considered setup.}  
    \label{figure:APsetup}  \vspace{-7mm}
\end{figure}

In this section, the numerical analysis is provided. There are $L=5$ transmit APs and $R=16$ receive APs with the height of $20$\,m and $50$\,m, respectively. The altitude of the target is $100$\,m, unless otherwise stated. The considered setup is depicted in Fig.~\ref{figure:APsetup} with $S=100$ sensing locations. The number of antenna elements per AP is $M=16$. There are $K=8$ UEs in the system, randomly located in a $500$\,m$\times 500$\,m area. The sensing area is $400$\,m$\times 400$\,m. Downlink transmit power at each AP is 1\,W and the minimum and maximum sensing blocklength are $50$ and $300$ symbols, respectively. The system is operating at $1.9$\,GHz with a bandwidth of $20$\,MHz and noise figure of $7$\,dB. False alarm probability threshold is set to $0.1$ and communication SINR threshold is $\gamma_c=10$\,dB, unless otherwise stated.
%\begin{table}[t]
%\caption{Simulation Parameters}  
%\label{tab:simpar}  
%\begin{tabular}{c|c}  
%\hline
%\toprule  
%Parameter & Value 
%\\  
%\midrule  
%TX-APs & $5$ \\
%RX-APs & $16$ \\
%UEs & $8$ \\
%Sensing points & $100$ \\
%Antennas per AP & $16$ \\
%False alarm probability & $0.1$ \\
%Number of setups with random UE locations &100 \\
%Number of setups for sensing & $1000$ \\
%Bandwidth &$20$ MHz \\
%Noise figure & $7$ dB  \\
%$\tau_{\text{s,min}}, \tau_{\text{s,max}}$  & %$50$, $300$\\
%Transmit power & $1$ W\\
%Area size & $400 \times 400$ $\text{m}^2$\\
%Carrier frequency& $1.9$ GHz\\
%Height difference between TX-AP and UE & $20$ %m\\
%Height difference between RX-AP and UE & $50$ m\\
%\bottomrule  
%\hline
%end{tabular}
%\end{table}
\enlargethispage{-0.1in}

%From the figure \ref{figure:APsetup}, the green square points represents TX APs, the red square points represents RX APs and blue round points represents possible locations for target drones. 

%During simulation, each epoch one location is selected and after all process is done for the previous location, the system will initialize all parameters, variables and iterative algorithm and start simulation for a next location.
\enlargethispage{-0.1in}

\begin{figure*}[t]
\begin{subfigure}{0.3\textwidth}
  \centering
  \includegraphics[trim={0mm 0mm 0mm 0mm},clip,width=1\linewidth]{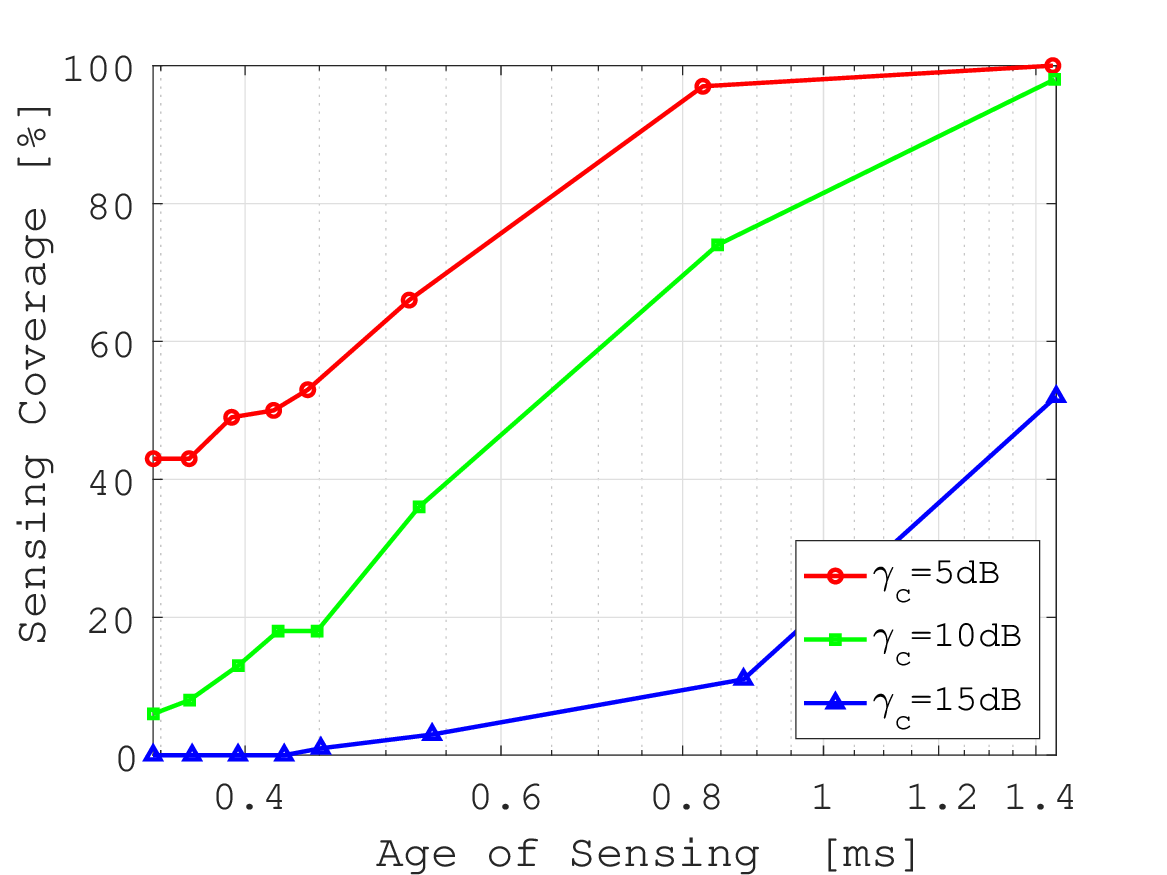} \caption{}
    \label{figure:time_vs_cov_gammac}
    \end{subfigure}
   \hfill
\begin{subfigure}{0.3\textwidth}
  \centering
  \includegraphics[trim={0mm 0mm 0mm 0mm},clip,width=1\linewidth]{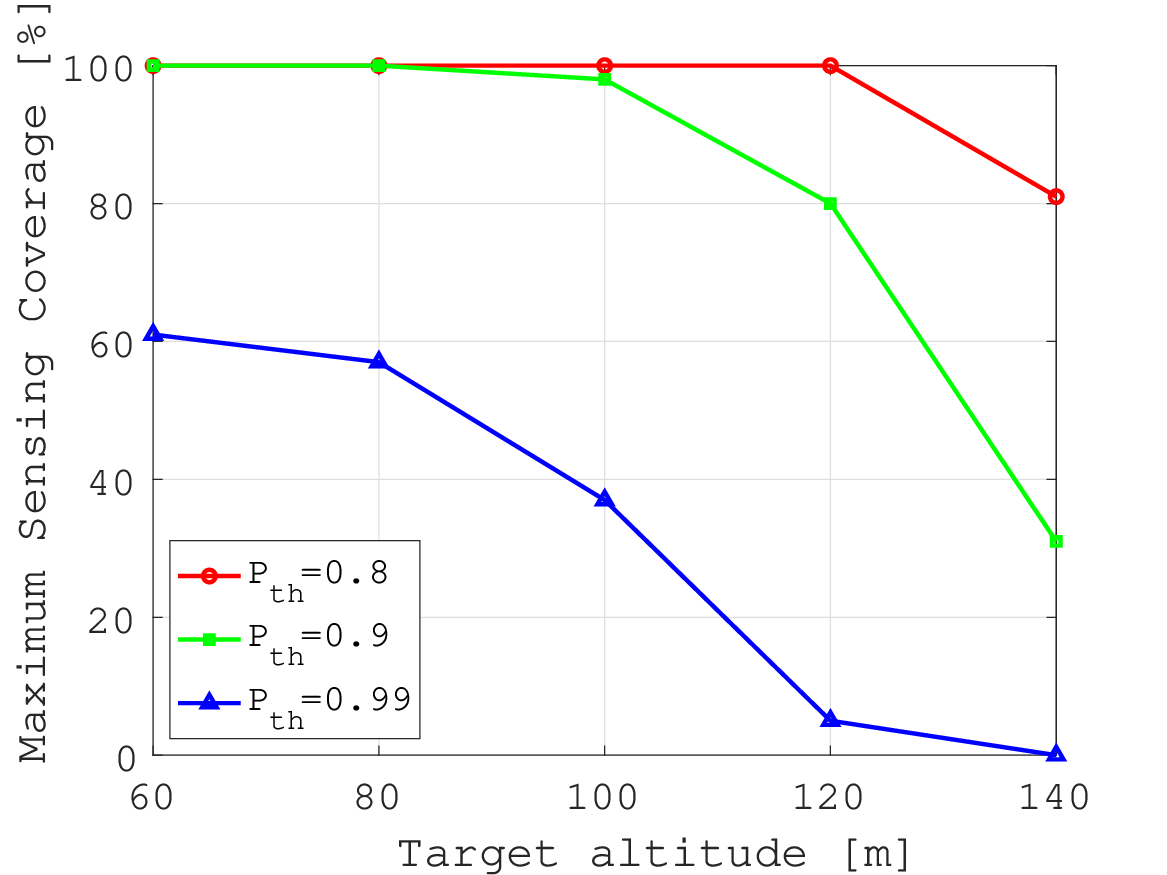} \caption{}
   \label{figure:height_cov_threshold}
    \end{subfigure}
    \hfill
    \begin{subfigure}{0.32\textwidth}
  \centering
  \includegraphics[trim={0mm 0mm 0mm 0mm},clip,width=1\linewidth]{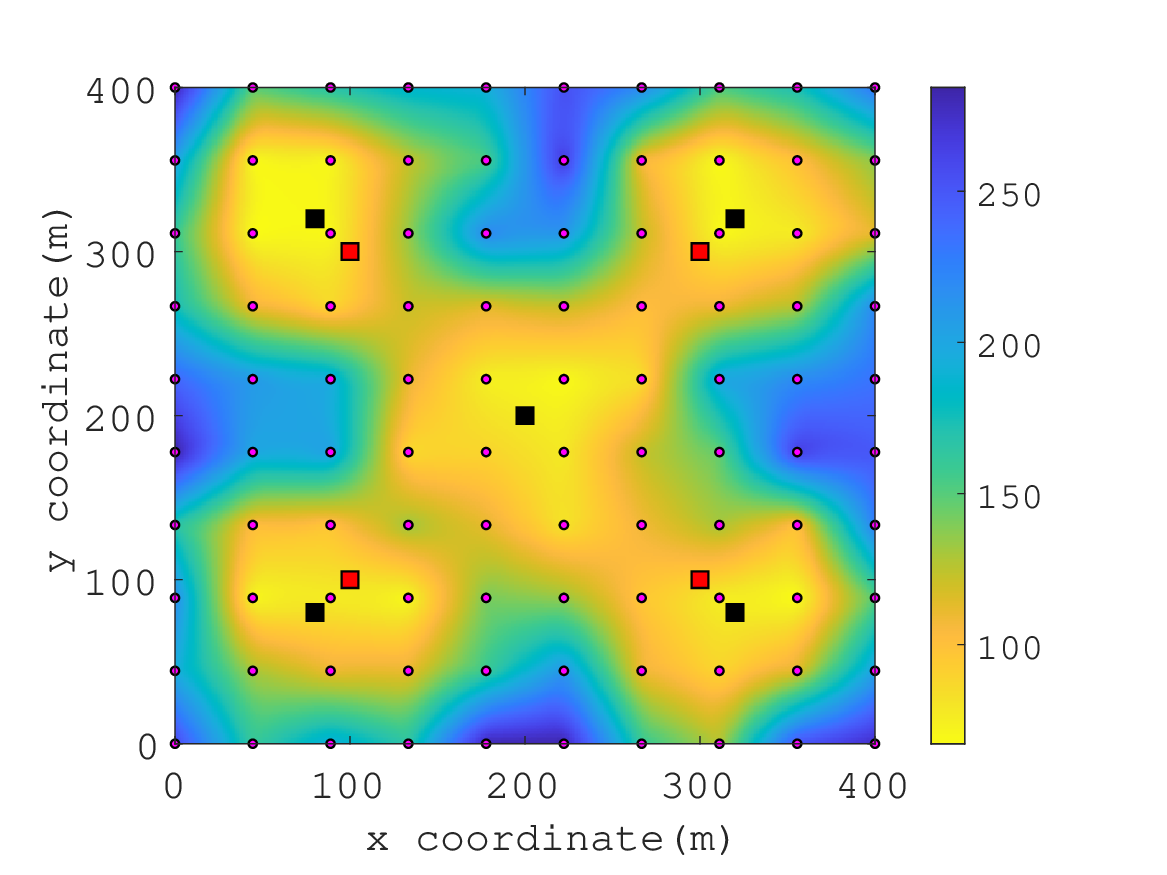} \caption{}
   \label{figure:tausmap_adapt}
    \end{subfigure}
    \vspace{-2mm}
   \caption{(a) Sensing coverage vs. total sensing time for $\gamma_c=5,10,15$\,dB and $P_{\rm th}=0.9$, (b) sensing coverage vs. target altitude for $P_{\rm th}=0.8,0.9, 0.99$, without adaptive weight selection algorithm and (c) minimum sensing blocklength with adaptive weight selection and $P_{\rm th}\!=0.9$.}
   \vspace{-6mm}
\end{figure*}

Fig.~\ref{figure:time_vs_cov_gammac} shows the relationship between sensing coverage and AoS for $S = 100$ sensing locations, under three different communication SINR threshold values. The detection probability threshold is $0.9$, and the curves are generated by adjusting the weights (\(\omega_0\) and \(\omega_1\)). 
As expected, sensing coverage increases with longer sensing durations, highlighting the trade-off between sensing coverage and timeliness of sensing information. Higher communication SINR thresholds reduce sensing coverage due to the need for higher power allocation for terrestrial UEs, leaving less power for sensing.
For \(\gamma_c = 5 \, \text{dB}\), sensing coverage starts at $40\%$ and rapidly approaches $100\%$, showing fast growth. For \(\gamma_c = 10 \, \text{dB}\), the initial coverage is lower (around $10\%$), but it accelerates notably between $0.4$ and $1$\,ms, eventually reaching approximately $100\%$ when total sensing time is larger than $1.2$\,ms. In contrast, \(\gamma_c = 15 \, \text{dB}\) demonstrates the slowest growth. Although coverage increases after $1$\,ms, it peaks at around $60\%$ reaching the maximum sensing blocklength. This behavior underscores the limitations of higher SINR thresholds on sensing coverage.
This illustrates how lower SINR thresholds enable faster sensing coverage, while higher SINR thresholds, despite providing better signal quality, require significantly longer sensing times to achieve similar coverage.

%\begin{figure}[t]
 %   \centering
 %   \includegraphics[width=0.65\columnwidth]{figures/cov_time_SINR.eps} 
 %   \caption{Sensing coverage vs. total sensing time for $\gamma_c=5,10,15$\,dB and $P_{\text{th}}=0.9$.}  
 %   \label{figure:time_vs_cov_gammac}\vspace{-4mm}  
%\end{figure}

The relationship between sensing coverage and target altitude is depicted in Fig.~\ref{figure:height_cov_threshold} for three different detection probability thresholds. Across all detection probability thresholds, sensing coverage decreases as altitude rises, indicating the challenge of maintaining coverage at higher altitudes. At $P_{\rm th}=0.8$, coverage starts at $100\%$ and decreases gradually, while for 
$P_{\rm th}=0.9$, coverage starts slightly lower and follows a similar decline. The most stringent threshold, $P_{\rm th}=0.99$, shows the steepest drop, starting at about $60\%$ coverage and quickly decreasing to near zero at higher altitudes, highlighting the trade-off between detection accuracy and sensing coverage.

Fig. \ref{figure:tausmap_adapt} presents the minimum sensing blocklength using the proposed adaptive weight selection algorithm, under the conditions of $\gamma_c=10$\,dB, target altitude of $100$\,m, and a detection probability threshold of $P_{\rm th}=0.9$. The x- and y-axis represent the simulation area, with the color gradient showing the optimized sensing blocklength required to meet the sensing performance criteria.
The yellow and green regions signify shorter blocklengths, reflecting better sensing performance in terms of AoS. A significant portion of the map appears yellow, forming a cross-like pattern between the transmit APs, where the sensing blocklength varies between $60$ and $100$ symbols in areas close to the APs. However, as the distance from the APs increases, the blocklength becomes considerably longer, reaching its maximum in the corner regions to meet the detection probability threshold. This demonstrates the spatial dependency of sensing performance on proximity to the transmit APs.

Table.~\ref{tab:tab2} compares the sensing coverage ($A_c$) and AoS  ($\Delta_{\mathrm{total}}$) using the proposed adaptive weight selection algorithm versus fixed weights for two points from Fig.~\ref{figure:time_vs_cov_gammac} when $P_{\rm th}=0.9$ and $\gamma_c=10$\,dB. With the proposed Algorithm 1, the total sensing time is $\Delta_{\text{total}}=0.757$\,ms, achieving a sensing coverage of $98\%$. In contrast, without the algorithm, the same sensing time of approximately $0.75$\,ms results in around $65\%$ coverage. To achieve the same $98\%$ coverage with fixed weights, the total sensing time increases significantly to $1.4$\,ms. The results demonstrate the advantage of the proposed algorithm, achieving up to a $45\%$ reduction in AoS compared to using fixed weights. 

\section{Conclusion}
   \vspace{-2mm}
In this work, we have proposed a cell-free massive MIMO-ISAC framework that aims unauthorized drone detection. The proposed framework optimizes jointly the power allocated to communication and sensing signals, and the sensing blocklength to balance between two main goals: (i) precision and (ii) timeliness, while guaranteeing communication requirements. We modeled the precision with the sensing coverage and timeliness with the AoS. The results demonstrate that the proposed method reduces AoS up to $45\%$, while maintaining over $98\%$ sensing coverage, demonstrating the efficient utilization of the resources. 

\begin{table}[]
\vspace{-2mm}
    \centering
    \caption{Comparison of sensing coverage and age of sensing.}
    \vspace{-2mm}
 \begin{tabular}{|c|c|c|}
    \hline
         \multirow{1.5}{*}{}&\multirow{1.5}{*}{$A_c$ [\%]}& \multirow{1.5}{*}{AoS [ms]}\\[1mm]
         \hline
         \multirow{1.5}{*}{With Algorithm 1}& \multirow{1.5}{*}{98}& \multirow{1.5}{*}{ 0.757}\\ [1mm]
         \hline\multirow{1.5}{*}{With fixed weights}& \multirow{1.5}{*}{$\approx 65$} & \multirow{1.5}{*}{0.75} \\[1mm]
         \hline\multirow{1.5}{*}{With fixed weights}& \multirow{1.5}{*}{$\approx 98$} & \multirow{1.5}{*}{1.4} \\[1mm]
         \hline
    \end{tabular}\label{tab:tab2}\vspace{-5mm}
\end{table}
\vspace{-4mm}

\vspace{-2.5mm}
\bibliographystyle{IEEEtran}
\bibliography{IEEEabrv,refs}
\end{document}